\documentclass{article}
\usepackage{spconf,amsmath,graphicx}
\usepackage{cite}
\usepackage{multirow}
\usepackage{url}
\usepackage[%
 setpagesize=false,%
 bookmarks=true,%
 bookmarksdepth=tocdepth,%
 bookmarksnumbered=true,%
 colorlinks=false,%
 pdftitle={},%
 pdfsubject={},%
 pdfauthor={},%
 pdfkeywords={}%
]{hyperref}

\title{\uppercase{Proposal of protocols for speech materials acquisition and presentation assisted by tools based on structured test signals}
}
\fourauthors
  {Hideki Kawahara\sthanks{Corresponding author, kawahara@wakayama-u.ac.jp}}
	{Wakayama University\\
	Center of Innovative Research and Liaison\\
	Wakayama, 640-8510 JAPAN}
  {Ken-Ichi Sakakibara\sthanks{kis@hoku-iryo-u.ac.jp}}
	{Health Sciences University of Hokkaido\\
	School of Rehabilitation Sciences\\
	Hokkaido, 061-0293 JAPAN}
  {Mitsunori Mizumachi\sthanks{mizumach@ecs.kyutech.ac.jp}}
	{Kyushu Institute of Technology\\
	Department of Electrical Engineering\\
	Kita-Kyushu, 804-8550 JAPAN}
  {Kohei Yatabe\sthanks{yatabe@go.tuat.ac.jp}}
	{Tokyo University of Agriculture and Technology\\
	Institute of Engineering\\
	Tokyo, 184-8588 JAPAN}

\begin{document}
\ninept
\maketitle
\begin{abstract}
We propose protocols for acquiring speech materials, making them reusable for future investigations, and presenting them for subjective experiments. We also provide means to evaluate existing speech materials' compatibility with target applications. We built these protocols and tools based on structured test signals and analysis methods, including a new family of the Time-Stretched Pulse (TSP). Over a billion times more powerful computational (including software development) resources than a half-century ago enabled these protocols and tools to be accessible to under-resourced environments.
\end{abstract}

\section{Introduction}
\label{sec:intro}



This paper is rather conceptual in its proposal to acquire more reusable speech materials and to present speech stimuli more reliably.
Due to the advancement of devices and media technologies, an enormous amount of digitized speech materials are accessible now.
Their acquisition conditions and physical characteristics are widely spread. 
In many cases, they are artificially processed and coded using a lossy scheme for storage and transmission.
Moreover, the speaking environment and situations strongly affect the speaker's behavior and produced speech attributes~\cite{Summers1988jasa,sodersten2005loud,Sierra2021frontier}.
It is crucial to disentangle these effects on speech to design applications using reusable speech materials and present speech outputs properly.

The following section describes the background and motivation of our proposal. 
Then, we introduce the organization of this article.


\section{Background}
\label{sec:background}

For scientific research and medical applications, demanding protocols for data acquisition are recommended~\cite{Svec2010splh,Patel2018speechHearing}.
Research tools for speech science and applications~\cite{eyben2010opensmile,boersma2011praat,Mcfee2015SciPy} are designed to derive acoustic attributes (such as intensity, spectral tilt, formant frequencies, and fundamental frequency, VOT: Voice Onset Time, shimmer, jitter, HNR: Harmonic to Noise Ratio, H1-H2: first and the second harmonic level difference, CPP: Cepstral Peak Prominence, and others) mainly from these well-qualified speech materials.
(Note that we use ``$f_\mathrm{o}$'' to represent the fundamental frequency instead of commonly used ``F0'' nor ``f0.''
Please refer to the discussions in the forum~\cite{titze2015jasaforum} for the background of this decision.)
The acoustic environment and situations strongly affect the values of these attributes.
Some attributes (for example, HNR and H1-H2) are susceptible while others (for example, $f_\mathrm{o}$ and CPP) are less.

Materials acquired under compatible conditions with the abovementioned demanding protocol are rare.
In everyday life, we are exposed to less-qualified speech materials without suffering from severe difficulties.
Current speech technologies, especially deep-learning-based ones, heavily depend on less-qualified materials (for example~\cite {Zhang2022icassp}).
Using less-qualified materials is inevitable and desirable because speech technologies have to be effective in conditions to which humans are exposed.
A recent review of objective assessment of synthetic speech suggested the acoustic attribute (in this case, reverberation) affects the quality of the synthesized speech~\cite{Cooper2024ast}.
A systematic method for providing objective indices to used speech materials helps develop better systems.

We sometimes observed reviewers of academic journals criticize the use of speech materials archived using lossy coding.
This criticism is not always relevant.
Depending on the target application, the coding scheme, and bitrate, sufficiently reliable derivation of some acoustic attributes from less qualified materials is possible.
Therefore, it is indispensable to bridge the gap between speech materials built for scientific research and materials we are exposed to and producing in everyday life by introducing objective evaluation and classification schemes of their quality (This issue is also discussed in ecological relevance of investigations ~\cite{keidser2020quest}).
The expansion of virtual environments and technologies made it possible to intervene in artificial acoustic environments and acquire their effects on vocalization data~\cite{hohmann2020virtual,Vorlaender2020acoustToday,Sierra2021frontier}.

We developed a set of procedures and tools for testing systems using structured test signals~\cite{Kawahara2023apsipa} enabled by two key technologies.
The first one is a new family of Time-Stretched-Pulse (TSP) called CAPRICEP (Cascaded All-Pass filters with RandomIzed CEnter frequencies and Phase polality~\cite{kawahara2021icassp}).
The second one is signal safeguarding~\cite{Kawahara2022ast}, which enables virtually all sound materials relevant for acoustic measurements.
We reformulated these technologies based on DFT's (DFT: Discrete Fourier Transform) orthogonal and periodic nature, taking advantage of efficient algorithms~\cite{Frigo1998icassp,Kumar2019cssp} and named this ``umbrella'' concept RAPHSODEE: Renewing impulse response measurement by post-Hoc Analysis using Periodic Stimulation with Orthogonal Decomposition for Extra Exploration.
Procedures based on this concept enable simultaneous measurement of multiple impulse responses using a structured, repetitive test signal consisting of multiple independent signals.
The responses provide the system's linear-time-invariant (LTI), signal-dependent time-invariant (SDTI), and random-and-time-varying (RTV) information.
We applied this concept RAPHSODEE in developing a tool for objectively measuring the involuntary voice $f_\mathrm{o}$ response to auditory stimulation~\cite{Kawahara2021apsipa,Kawahara2023smac} and a Lombard effect measurement tool based on a real-time modification of acoustic environment including auditory feedback path~\cite{Kawahara2023smac}.
We modified the voice $f_\mathrm{o}$ response tool for the objective evaluation method for ``pitch'' extractors~\cite{Kawahara2022isFoMes}.

We realized we have tools to objectively evaluate and classify speech acquisition, archiving, processing, and presentation.
These motivated us to develop and propose protocols to make speech materials reusable and make presentation proper.
Our proposed protocols cover under-resourced situations by combining these technologies with about a billion times more powerful computation than a half-century ago~\cite{Leiserson2020science}.

\section{Assessment of acoustic conditions}
\label{sec:acquisition}
\begin{figure}[tp]
\centering
\includegraphics[bb=0 0 374 179, width=8.5cm]{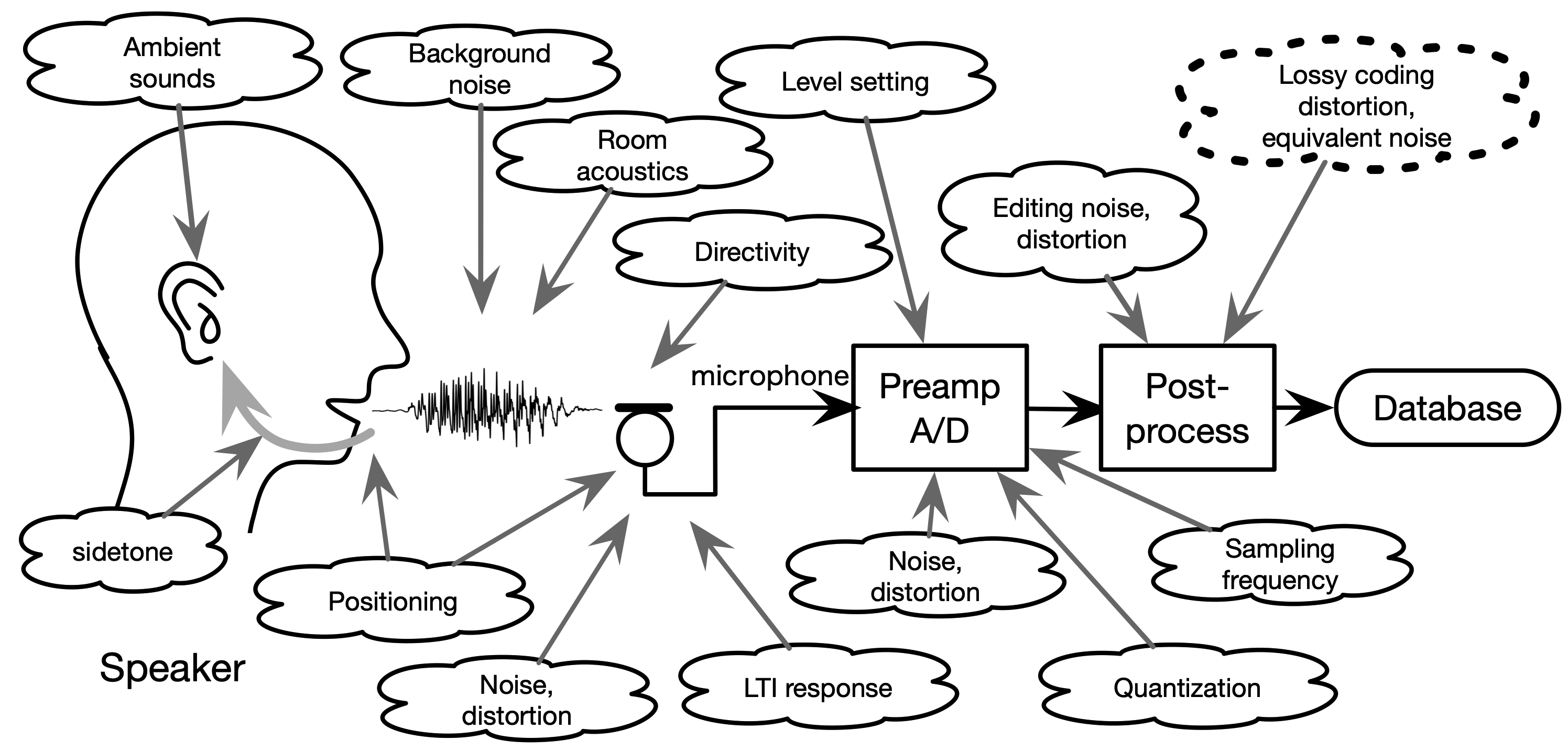}
\caption{Contributing factors affecting speech material acquisition (adopted from~\cite{Kawahara2023smac}).\label{fig:aquisitionCondition}}
\end{figure}

Figure~\ref{fig:aquisitionCondition} illustrates the contributing factors involved in acquiring speech materials.
Providing relevant information in the figure to the acquired speech materials and obeying the recommended protocol~\cite{Patel2018speechHearing} enables the speech materials to be reused for further scientific research.

\begin{figure}[tp]
\centering
\includegraphics[bb=0 0 397 148, width=8.5cm]{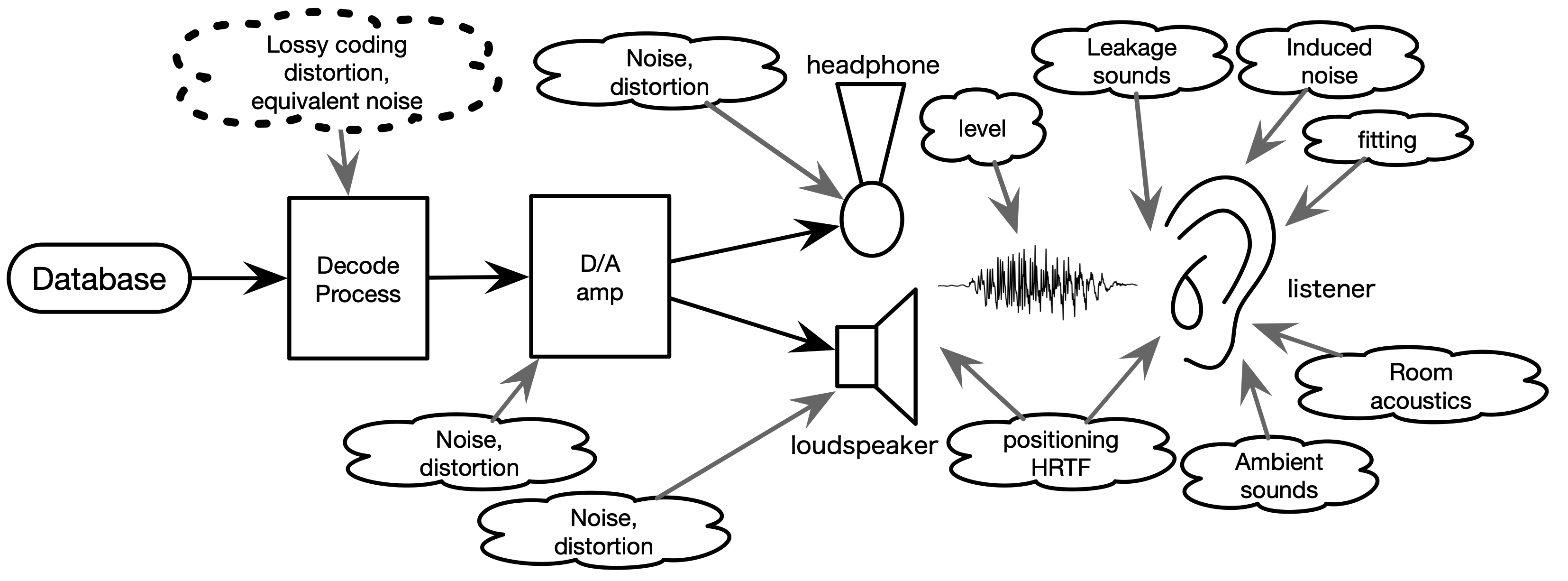}
\caption{Contributing factors affecting speech presentation.\label{fig:presentationSchem}}
\end{figure}

Figure~\ref{fig:presentationSchem} shows contributing factors in presenting speech materials.
While objective measurement of the transducers and subjective test recommendations exist~\cite{MUSHRA,ITU-TP57,ITUp58}, actual presented and interfering sounds (leakage of ambient and induced sounds) differ significantly in everyday life situations.
Furthermore, cloud sourcing acquisition and tests introduce additional discrepancies.
Refer to the above-mentioned recommendations, recommendations for scientific research~\cite{Svec2010splh,Patel2018speechHearing}, and textbooks (for example~\cite{beranek2019acoustics}).

\subsection{Relevant attributes}
\label{sec:attributes}
This section discusses attributes relevant to assessment.
The following attributes are informative as the metadata to the acquisition and the presentation.

\subsubsection{Acquisition related attributes}
\label{sec:Acquisition}
The first list is for the acquisition of speech materials.
\begin{description}
\item[Sound pressure level: ] Voice production is a physiological process. 
Sound pressure level at a physically defined reference point is essential to asses voice conditions.
\item[Background noise (RTV): ]  This consists of ambient sounds, noise in sound acquisition equipment (microphone, pre-amplifier), and noise due to A/D conversion and coding (especially lossy coding; this is signal dependent).
\item[Distortion (LTI): ]  LTI distortion is the deviation from frequency-independent gain and phase. The most contributing factor is the microphone frequency response. Positioning (distance and orientation of microphone and speaker's orientation) and the directivity pattern of the microphone also contribute.
\item[Distortion (SDTI): ]  Exceedingly high-level microphone input and pre-amplifier input introduces signal-dependent and time-invariant distortion due to their non-linearity. Lossy coding also introduces signal-dependent distortion.
\item[Direct sound to indirect sound: ] In everyday life, the acquired speech consists of the directly arrived sound and sounds reflected by surrounding objects and reverberation.
\item[Post processing: ] Sound materials available on media (such as broadcasting, Internet, audiobook, CD, and streaming) are post-processed and edited. These processes irreversibly alter the original recording.
\end{description}

\subsubsection{Presentation related attributes}
\label{sec:Presentation}
The second list is for the presentation of speech materials.
\begin{description}
\item[Sound pressure level: ] Huma auditory system is highly-nonlinear. Its behavior significantly depends on the presented sound pressure level.
\item[Background noise (RTV): ]  This consists of ambient sounds, noise in sound reproduction equipment (loudspeaker, headphones, and amplifier), and noise due to decoding of lossy coding materials. In addition to these, induced sounds of scratching cable and biological noise (blood flow, pulsive muscle noise) for headphones. Loudspeakers introduce signal-dependent random noise (turbulent noise due to airflow induced by the diaphragm movement and high-frequency phase modulation due to the Doppler effects of breeze.). High power dissipation in loudspeakers also introduces temporal variations of the system.
\item[Distortion (LTI): ]  The most contributing factor is the frequency responses of the loudspeaker and headphones. For loudspeakers, the positioning of the loudspeaker and the listeners, together with the room acoustics LTI distortion, significantly varies. For headphones, the fitting condition has significant effects.
\item[Distortion (SDTI): ]  Effects of non-linearity are more severe in loudspeakers than headphones.
\end{description}

We are developing tools for measuring these attributes based on the RHAPSODEE concept.
The following section introduces measurement methods based on it.

\section{RHAPSODEE and measurement}
\label{sec:Rhapsodeemeasurement}
This section outlines the underlying principles and tools of simultaneous measurement of multiple impulse responses and attributes derivation from the responses.
For technical details and descriptions of tools, refer to~\cite{Kawahara2023apsipa}.
We refined the procedures and tools since~\cite{kawahara2021icassp,Kawahara2023apsipa}.
We remark them in the following descriptions.

\subsection{Impulse response measurements}
\label{sec:measurement}
\begin{figure}[tp]
\centering
\includegraphics[bb=0 0 1839 920, width=8.5cm]{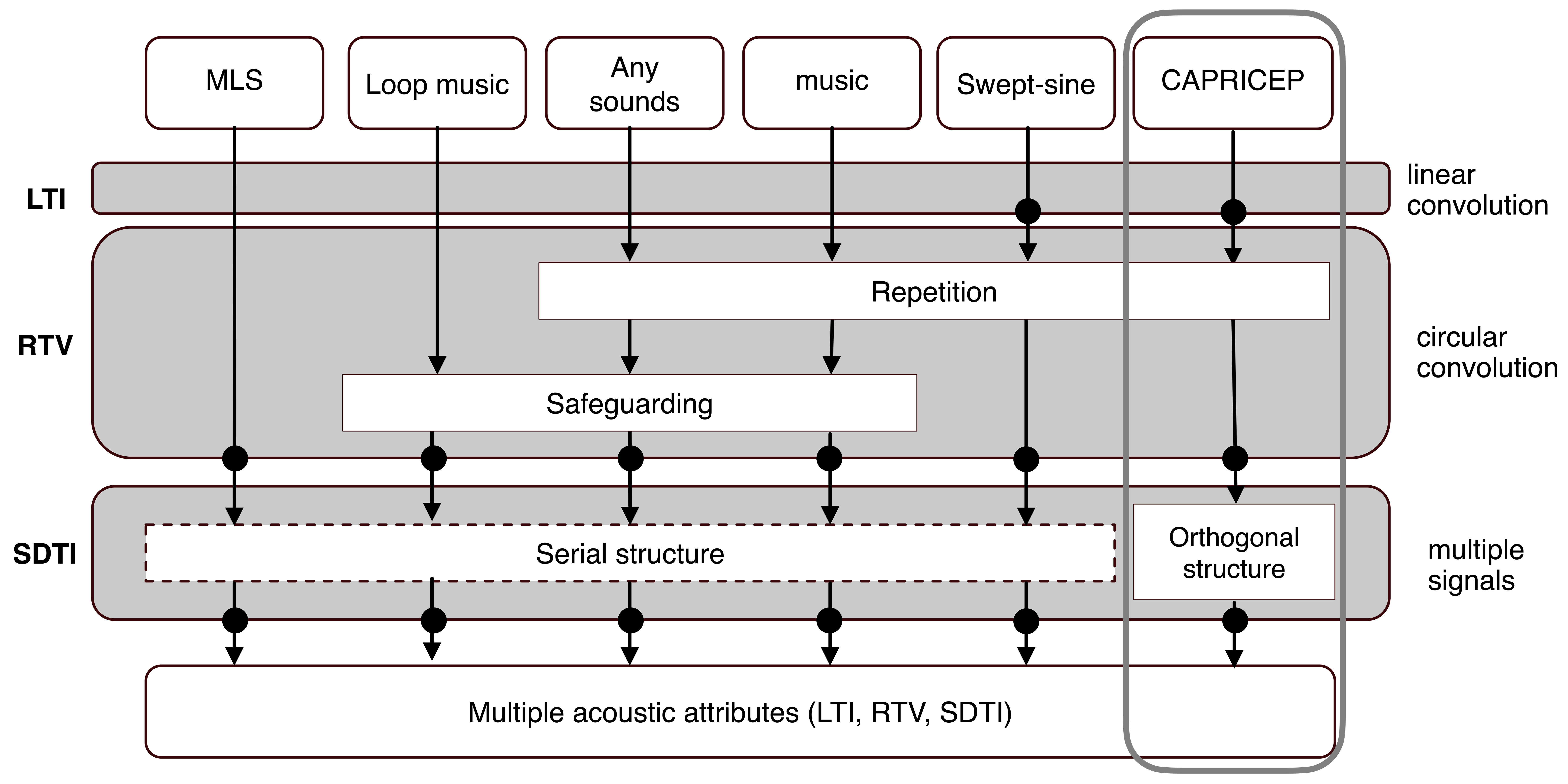}
\caption{Schematic diagram of RHAPSODEE. This diagram is a refined version of Fig.1 in~\cite{Kawahara2023apsipa}. The rightmost gray frame represents the test signal for the simultaneous measurement. In the figure, ``MLS: Maximum Length Sequence'' and ``Swept-sine'' are commonly used test signals for impulse response measurements~\cite{Aoshima1981jasa,Rife1989aes}. The term ``TSP: Time Stretched Pulse'' represents them, and our CAPRICEP~\cite{kawahara2021icassp} is a new family member of TSP. Black dots indicate where output is available. \label{fig:RHAPSODEEc}}
\end{figure}

Figure~\ref{fig:RHAPSODEEc} illustrates the concept of RHAPSODEE.
Let discrete Fourier transform of the input signal $x[n] \xrightarrow{\mathcal{F}}  X[k]$ and the output signal $y[n] \xrightarrow{\mathcal{F}}Y[k]$ of the target system are known.
Then, as long as the system is LTI, the transfer function $H[k]$ and the impulse response $H[k] \xrightarrow{\mathcal{F}^{-1}} h[n]$ are derived as the division and its inverse Fourier transform.
$k$ represents the discrete frequency and $n$ represents the discrete time.
The symbols $\xrightarrow{\mathcal{F}}$ and $\xrightarrow{\mathcal{F}^{-1}}$ represents the discrete Fourier transform and its inverse.

\subsubsection{Linear convolution for LTI: top gray block of Fig.~\ref{fig:RHAPSODEEc}}
\label{sec:liearconv}
Swept-sine and unit-CAPRICEP are TSP and are finite in length.
Using the buffer length that is longer than the sum of TSP's length and the (effective) length of the impulse response of the target system, circular convolution using DFT is identical to linear convolution.
The top gray block with two black terminals represents this impulse response measurement.
(The original CAPRICEP~\cite{kawahara2021icassp} used the phase function of IIR all-pass filters. We revised it using Gaussian-shaped group delays to design the all-pass phase functions. This revision improved the temporal localization of CAPTICEP impulse response~\cite{Kawahara2023apsipa}.)

\subsubsection{Circular convolution for RTV: middle gray block of Fig.~\ref{fig:RHAPSODEEc}}
\label{sec:circConv}
The repetitive presentation of a signal yields a periodic signal.
When the repetition period is longer than the target system's impulse response length, the output signal becomes a periodic signal after the second repetition.
Then, after the second repetition, the ratio of the DFTs of the aligned input and output signals' segment with the length of the repetition period provides the transfer function and the impulse response of the target system.
Note that the derived impulse response is identical, irrespective to the segment locations, as long as they are after the second repetition.

There is a caveat.
Fourier transform of the signals other than TSPs have elements with small absolute values.
They introduce significant susceptibility to measurement noise.
We introduced signal safeguarding~\cite{Kawahara2022ast} by flooring the denominator's minimum absolute value to make such signals relevant for impulse response measurement.
(The original safeguarding~\cite{Kawahara2022ast} used a frequency-independent thresholding. We revised to use spectral shaping in thresholding~\cite{Kawahara2023apsipa}.)

The derived impulse responses from different locations are different in the actual measurement.
This is because of background noise and the system's temporal variation, the RTV component.
Using segments of several repetition cycles provides estimates of the RTV component.
The middle gray block with six black terminals represents repetition-based RTV derivation.

\subsubsection{Multiple siganls for SDTI: bottom gray block of Fig.~\ref{fig:RHAPSODEEc}}
\label{sec:sdtiMesurement}
Averaging impulse responses derived from each repetition yields better estimations of LTI.
When the system is strictly LTI, the averaged impulse responses are the same regardless of input test signals.
However, real-world systems are non-linear. 
Therefore, the impulse responses derived using different test signals differ.
Using impulse responses derived from different test signals provides estimates of the SDTI component.
It requires serial measurements using different test signals.

The close-to-orthogonal nature of different CAPRICEP signals enables simultaneous measurement of multiple impulse responses to different test signals.
We introduced systematic allocation of different CAPRICEP signals using the Walsh-Hadamard matrix to prepare the structured test signal~\cite{Kawahara2023apsipa}.
After several repetitions of this structured test signal, it turns into a periodic signal.
It also is a TSP.
We developed procedures to derive LTI, RTV, and SDTV simultaneously using a fast implementation of DTF~\cite{Frigo1998icassp,Kumar2019cssp,Kawahara2023apsipa}.
(Again, this implementation is substantially more simple and more efficient than the original article~\cite{kawahara2021icassp}.)
The bottom gray block with six black terminals represents multiple measurement-based SDTI derivation.

\subsection{Measurement of acoustic attributes}
\label{sec:acousticAttributeMeasre}
Once an impulse response (LTI information) is available, it provides acoustic attributes that contribute to the quality of the acquisition and presentation.
For example, it provides a bendwise signal-to-noise ratio by comparing the RTV component and the speech spectrum.
It significantly contributes to the measurement of HNR.
Power decay of bandpass outputs provides reverberation radius (distance at which indirect sound and direct sound equal).
It is recommended that the microphone distance should be less than half of the reverberation radius~\cite{Patel2018speechHearing}.
Providing this information as metadata to speech materials is essential.
Because the indirect sound (reverberation) level of the training speech materials contributed to synthesized sound quality~\cite{Cooper2024ast}.

\begin{figure}[tp]
\centering
\includegraphics[bb=0 0 997 265, width=8.5cm]{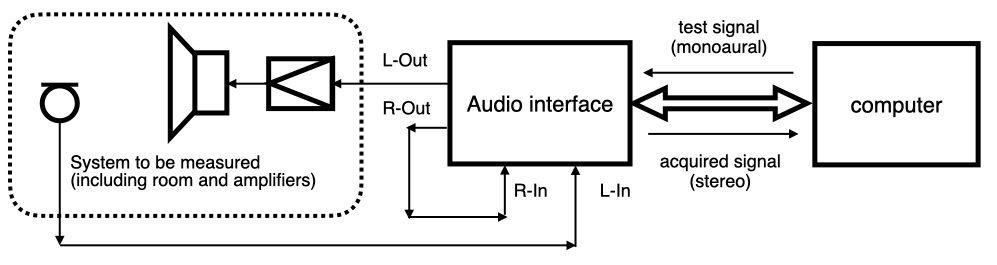}
\vspace{-3mm}
\caption{Assessment setup for input system test. Connection from R-out to R-in is not compulsory. \label{fig:measuringSystemOC}}
\end{figure}

Figure~\ref{fig:measuringSystemOC} shows a setting for measuring acoustic attributes of speech material acquisition.
This setting replaces the human speaker with a loudspeaker.
Reproducing the structured test signal and recording the sound and the test signal provides the information for LTI, RTV, and SDTI.

\begin{figure}[tp]
\centering
\includegraphics[bb=0 0 1064 777, width=8.5cm]{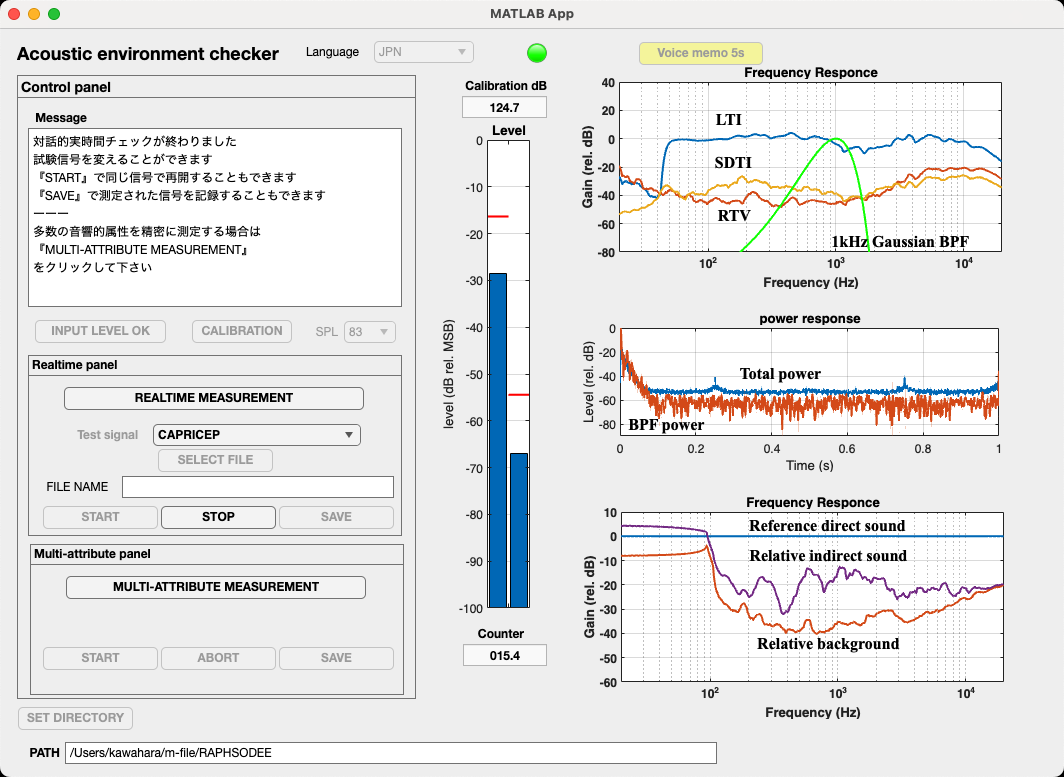}
\caption{GUI snapshot of the interactive tool for acoustic condition assessment. Times-Roman notes explain lines.\label{fig:testTool}}
\end{figure}

\begin{figure}[tb]
\begin{center}
\includegraphics[bb=0 0 1268 840, width=8.5cm]{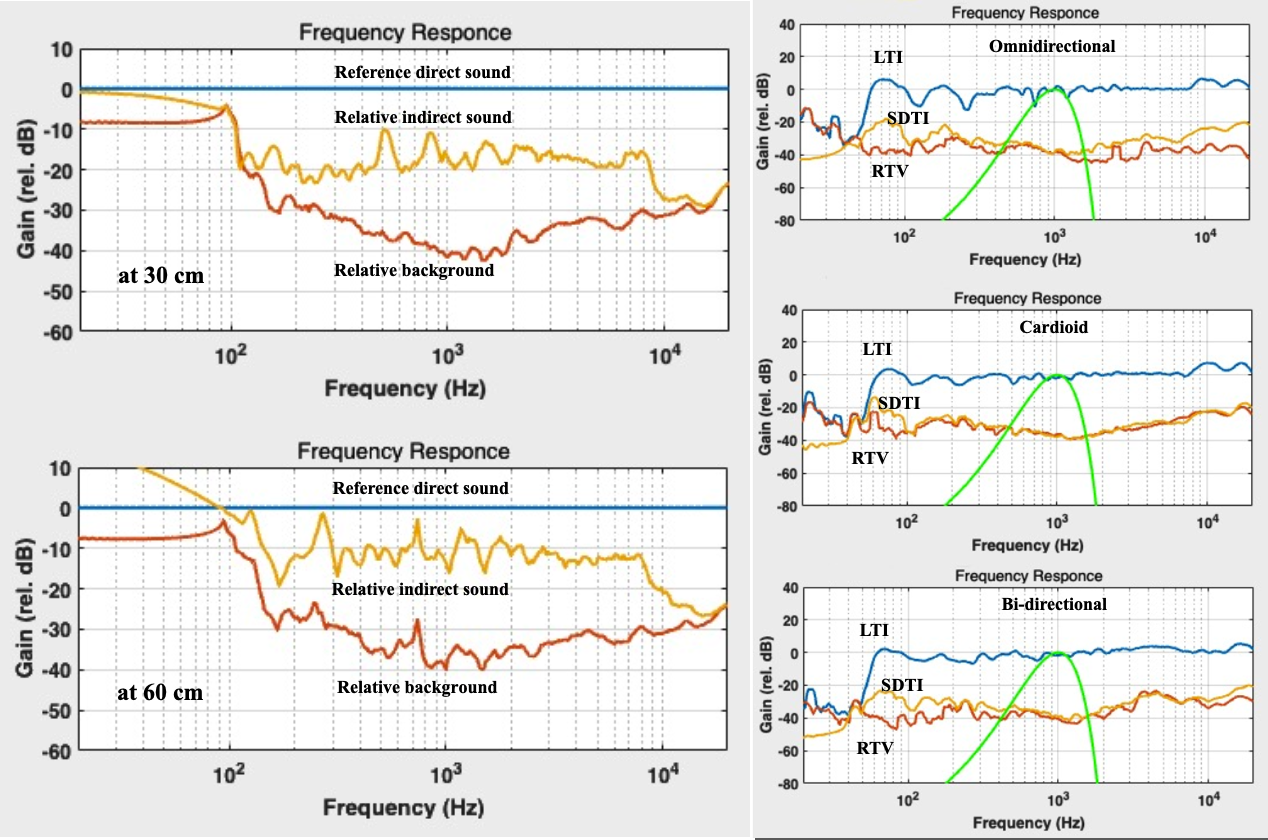}
\vspace{-3mm}
\caption{(Left two plots:) Direct vs indirect ratio and RTV ratio measured at 30~cm (upper plot) and 60~cm (lower plot).
(Right three plots:) Frequency response dependency on the microphone directivity pattern. From top to bottom, omnidirectional, cardioid, and bi-directional patterns. Refer to Times-Roman notes for the explanation.}
\label{fig:distanceEffect}
\end{center}
\vspace{-3mm}
\end{figure}

\subsection{Tools for acoustic measurement}
\label{sec:Tools}
We developed tools for acoustic measurement based on RHAPSODEE~\cite{Kawahara2023apsipa}.
Figure~\ref{fig:testTool} shows a snapshot of the GUI of the tool.
The tool was developed using Appdesigner SDE of MATLAB (R2024 Update 6) on a MacBook Pro M1 Max (Apple, 64GB memory, and 1TB SSD).
This section provides an overview of the tool and measurement examples.
Refer to~\cite{kawahara2023asjAA} for detailed descriptions about step-by-step procedure of measurement.
We also prepared a YouTube channel for this tool and previous versions.\\ {\tiny (\url{https://www.youtube.com/playlist?list=PLqr7NXdG4BylP8UEHn8lsSF4zK4KE3kw_})}

The left panel is for the tool's control functions. 
The top subpanel displays messages indicating the tool's current state and the user's required operation.
(Japanese and English messages are selectable.)
The middle subpanel is for real-time measurement, and the bottom is for detailed off-line measurement.
Just under the message area is for sound pressure level calibration of the input system.

The center bar graph is for the input level monitor.
The three plots on the right-hand display measurement results.
The top displays the frequency response of LTI, RTV, and SDTI.
It also indicates the frequency response of the band-pass filter for power decay measurement.
The middle plot displays the power decay of the whole frequency range and the band-pass filtered impulse response.
The bottom plot displays the relative level of the indirect sound and the background noise, RTV.
The horizontal line represents the normalized direct sound level.
We added explanatory notes to the GUI snapshot using Times-Roman font.

\subsubsection{Examples}
\label{sec:example}
We tested several conditions of microphone setting and directivity patterns.
Figure~\ref{fig:distanceEffect} shows the distance effects of microphone placement.
In this example, we used a large diaphragm condenser microphone with selectable directivity (omni, bi, and cardioid pattern), RODE NT2-A.
The audio interface was Zoom F3, using a 44100~Hz sampling rate and 32-bit floating number format.
The recording was conducted in a Japanese room with an 18 cubic meter volume.
Three lines in each plot represent, from top to bottom, the direct sound LTI (reference), indirect sound LTI, and RTV.
This example indicates that at 60~cm, the effects of the indirect sound are not negligible.

Three plots on the right-hand side of Fig.~\ref{fig:distanceEffect} show the effect of the directivity pattern on the LTI response.
Note the difference between the blue lines in these plots (the darkest lines in BW print).
The plots indicate that a bi-directional pattern provides minimum LTI distortion in everyday life.
We also tested the effects of sound shields on LTI distortion.
The best condition is the combination of the cardioid-directivity facing the null point to the sound shield.
The combination of the omnidirectional directivity introduced significant distortion.
Real-time displays of the LTI response and the direct-to-indirect ratio are helpful for the microphone setting adjustment.

\begin{table*}[tp]
\caption{A tentative plan for sound material classification}
\begin{center}
\begin{tabular}{c|lll}\hline \\
Class & Summary & Acquisition and Annotation & reusability \\ \hline
\multirow{5}{*}{1} & background noise: minimum & records background characteristics & appropriate for analyzing \\
  & reverberation: minimum & (records acoustic field information) & all acoustic attributes  \\
  & (ex: SNR $>$ 30~dB) &  Using high-quality microphone  &   \\
  &  refer~\cite{Svec2010splh,Patel2018speechHearing} &  precise calibration of sound pressure level  &  \\
  &   &  detailed annotation of recording session &  \\ \hline
\multirow{5}{*}{2} & background noise: minimum1 & records background characteristics &  appropriate for analyzing \\
  & reverberation: minimum &  (records acoustic field information)  & many acoustic attributes  \\
  &   &  Using high-quality microphone &   \\
  &   &  sufficient calibration of sound pressure level &  \\
  &   &  relevant annotation of recording sessio &  \\ \hline
\multirow{4}{*}{3} & High quality recording & records background characteristics & appropriate for analyzing \\
  & in fields and everyday-life  & Using sufficient quality microphone & some acoustic attributes  \\
  & situations  &  calibration of sound pressure level &   \\
  &   &  annotation of recording sessio &  \\ \hline
\multirow{3}{*}{4} & recording in fields and & recording of a simple test signal & acoustic attributes analysis \\
  & everyday-life situations & (discussion needed)  & can be possible with preprocessing  \\
  &   &    &  usable for training  \\
\hline
\end{tabular}
\vspace{-2mm}
\end{center}
\label{tbl:class}
\end{table*}%

\subsection{Simple test signal for annotation of recording condition}
\label{sec:simplesignal}
The introduced tools are relevant for recording in laboratories.
However, using these tools is impractical for recording in the field and in everyday situations.
We designed a simple test signal that sequences the structured test signal, followed by three seconds of silence and a calibration signal.
After completing the speech material acquisition, playing back this test signal using, for example, a smartphone or a powered speaker placed at the speaker's (informant's) position and recording it as a supplemental file enables analysis of recording conditions afterward.
It is essential to leave note of the sound pressure level of the calibration tone at the microphone using a sound level meter.
Smartphone applications are sufficient in this case~\cite{Faber2017acoustictoday}.

\section{Classification and protocols}
\label{sec:classification}
Table~\ref{tbl:class} suggests a tentative plan of speech material classification.
The RHAPSODEE-based tools measure acoustic attributes needed for metadata.
For class 4 materials, adding a recording of the simple test signal improves their reusability.
We will discuss the classification and desirable tools and protocols for data acquisition with researchers, practitioners, and cloud-sourcing participants.


\section{conclusion}
\label{sec:conclusion}
We introduced protocols and tools for speech material acquisition and presentation.
We built these protocols and tools based on structured test signals and analysis method, including a new family of the Time-Stretched-Pulse (TSP).
The combination of these tools with recent advancements in computational power and audio equipment enabled these protocols and tools to be accessible to under-resourced environments.
We open-sourced these tools and materials~\cite{kawahara2020gitHk}.

\section{Acknowledgement}
\label{sec:ack}

This work was supported by JSPS KAKENHI Grant Numbers  JP20H00291, JP21H00497, JP21H01596, and JP21K19794. 

{\small
\bibliographystyle{IEEEbib}
\bibliography{kawahara2024ococosda}
}

\end{document}